# Accurate analysis for harmonic Hall voltage measurement for spin-orbit torques

Seok Jin Yun,[1] Eun-Sang Park,[2] Kyung-Jin Lee,[1,2] and Sang Ho Lim[1,*]

[1]*Department of Materials Science and Engineering, Korea University, Seoul 02841, Korea*
[2]*KU-KIST Graduate School of Converging Science and Technology, Korea University, Seoul 02841, Korea*

An accurate method is developed to extract the spin-orbit effective fields through analysis of the results of harmonic Hall voltage measurements by deriving detailed analytical equations, in which both the *z*-component of the applied magnetic field and the second-order perpendicular magnetic anisotropy are taken into account. The method is tested by analyzing the results of a macrospin simulation. The spin-orbit effective fields extracted from the analysis are found to be in excellent agreement with the input spin-orbit effective fields used for the macrospin simulation over the entire range of the polar magnetization angle and a wide range (0–2) of the ratio of the planar to the anomalous Hall voltage considered in this study. The accuracy of the proposed method is demonstrated more clearly via a systematic study involving a comparison of its results with those of the conventional analytical method.

Running title: accurate analysis for harmonic Hall voltage

[*]Corresponding author. Name: Sang Ho Lim; Postal address: Department of Materials Science and Engineering, Korea University, 145 Anam-ro, Seongbuk-gu, Seoul, Korea; Tel.: +82-2-3290-3285; fax: +82-2-928-3584; e-mail: sangholim@korea.ac.kr

# I INTRODUCTION

It has recently been found that in-plane currents in a nonmagnetic (NM)/ferromagnetic (FM) bilayer nanostructure can generate a torque due to the spin-orbit (SO) coupling, known as the spin-orbit torque (SOT), which is sufficient enough to reverse the magnetization in the FM layer.[1] Numerous studies have been conducted to identify the principal mechanism of the SOT as being either the spin Hall effect (SHE) in the NM layer[2-4] or the interfacial spin-orbit coupling (ISOC)—frequently referred to as the Rashba effect—at the NM/FM interface.[5-12] In a system in which the NM/FM interface is perpendicular to the $z$-axis and the in-plane currents flow along the $x$-axis, spin currents polarized along the $y$-axis are generated in the system on the basis of the SHE induced by a bulk SO coupling in the NM layer. The spin currents are injected into the adjacent FM layer, thus causing transfer of a torque to the magnetization of the FM layer. The SHE-induced SOT generates a strong damping-like torque ($\boldsymbol{T}_{DL} \propto \boldsymbol{m} \times \boldsymbol{m} \times y$) but a weak field-like torque ($\boldsymbol{T}_{FL} \propto \boldsymbol{m} \times y$).[13,14] Theoretically, the strength of the SHE-induced SOT is known to be independent of the magnetization direction of the FM layer. In the case of the ISOC-induced SOT, spins polarized along the $y$-axis accumulate owing to the broken inversion symmetry at the NM/FM interface. Direct exchange coupling between the magnetization of the FM layer and the accumulated spins generates a strong $\boldsymbol{T}_{FL}$ but a weak $\boldsymbol{T}_{DL}$.[7,15-18] Unlike the strength of the SHE-induced SOT, that of the ISOC-induced SOT is known to depend on the magnetization direction of the FM layer.[19-21] In these two cases, both the SHE and the ISOC qualitatively induce the same torque on the FM layer. In order to identify the dominant mechanism of the SOT, a quantitative analysis of the values of $T_{DL}$ and $T_{FL}$ over a wide range of magnetization angles is required.[19–21]

The harmonic Hall voltage measurement method is one of the useful approaches for quantifying the effective fields of $T_{DL}$ and $T_{FL}$ originating from the SOT.[22] This method is particularly suited for identifying the angular dependence of the SOT acting on the FM layer with perpendicular magnetization.[19,20] Several corrections are required to be made for an accurate analysis of measured results, including for the following: the planar Hall effect (PHE),[20,23] the out-of-plane component of the external magnetic field,[20] and the anomalous Nernst effect (ANE).[20,24] In the harmonic Hall voltage measurement, the second harmonic voltage ($V^{2\omega}$) consists of two major components: anomalous and planar Hall voltages (denoted as $V_{AHE}$ and $V_{PHE}$, respectively).[20,23] When an external magnetic field ($H_{ext}$) is applied along the longitudinal ($x$) direction, the $V^{2\omega}$ values resulting from the AHE and PHE are proportional to $T_{DL}$ and $T_{FL}$, respectively. Under a transverse ($y$) $H_{ext}$, however, these values are proportional to $T_{FL}$ and $T_{DL}$, respectively. This necessitates the use of an analytical solution based on Cramer's rule in order to separate $T_{FL}$ and $T_{DL}$.[23] This analytical solution has been successful only for a system with $V_{PHE} < V_{AHE}$. For a system with $V_{PHE} > V_{AHE}$, such as the W/CoFeB/MgO trilayer structure,[25] a divergence occurs in the solution, making it extremely difficult to analyze the measurement results.

This problem can be overcome by making some necessary corrections in the analysis of the measurement results, including that for the out-of-plane component of $H_{ext}$. Since coherent magnetization rotation is an important requirement in the analysis of harmonic Hall voltage measurement results, $H_{ext}$ is usually applied along the direction tilted slightly ($4°−15°$) from the basal plane (*x*-*y* plane).[20] Under this condition, the *z*-component of $H_{ext}$ has a nonzero value, although it has been neglected thus far just to simplify the analysis. This assumption is reasonable in the low-$H_{ext}$ range, where the magnetization direction is close to the *z*-axis and the perpendicular magnetic anisotropy (PMA) field is consequently dominant over the *z*-component of $H_{ext}$.[23] The simplifying assumption, however, is no longer valid in the high-$H_{ext}$ range, where the magnetization direction is considerably deviated from the *z*-axis with a resultant reduction in the PMA field, as a result of which it loses its dominance over the *z*-component of $H_{ext}$. Several attempts have been made in the past to include the *z*-component of $H_{ext}$, which is obtained by solving equilibrium torque equations repetitively until a desired convergence is achieved (the recursive method).[19,20] However, this method is quite complicated; furthermore, it has not been validated for systems with $V_{PHE} > V_{AHE}$. Unwanted voltages, which originate from thermoelectric effects such as the ANE, should also be eliminated from the harmonic signals. Although several methods have been proposed for this purpose,[20,24] erasing all the artificial signals remains difficult. Another important issue that needs to be addressed is the inclusion of the second-order PMA—which has not been taken into account thus far—in the analysis of harmonic Hall voltage measurement results. The inclusion of the second-order PMA is considered to be of great importance, because many PMA materials exhibit the second-order PMA, with its strength being comparable to that of the first-order PMA in many cases.[26,27]

In the present study, two corrections—one for the $z$-component of $\boldsymbol{H}_\text{ext}$ and the other for the second-order PMA—are considered in the analysis of the harmonic Hall voltage measurement results. All the related analytical equations are described. Both the conventional and the refined analytical methods are used to analyze the results of a macrospin simulation, which plays a role of mimicking the harmonic Hall voltage measurement by numerically solving the Landau–Lifshitz–Gilbert equation.[28,29] The accuracy of these two analytical methods is tested by comparing the input SO effective fields used for the macrospin simulation with those calculated by the analytical methods. To test the refined analytical method critically, systems are examined over a wide ratio $R$, which is defined as $V_\text{PHE}/V_\text{AHE}$. A similar comparative study was also performed that involves analyzing the experimental results of harmonic Hall voltage measurements for a Pt/Co/MgO structure.

## MATERIALS AND METHODS

### A. Analytical solutions for conventional approach

When an in-plane AC current with frequency $\omega$ ($I_\text{AC} = I_0 \sin\omega t$) is applied to an NM/FM bilayer structure, the angle between the $z$-axis and the magnetization of the FM layer ($\theta_M$) and that between the $x$-axis and the orthographic projection of the magnetization on the $x$-$y$ plane ($\varphi_M$) oscillate as $\theta_M(t) = \theta_M{}^\circ + \Delta\theta_M \sin\omega t$ and $\varphi_M(t) = \varphi_M{}^\circ + \Delta\varphi_M \sin\omega t$. Here, the superscript $^\circ$ and symbol $\Delta$ denote the value in the absence of $I_\text{AC}$ and the amplitude of the related angles, respectively. The total energy equation for the NM/FM bilayer structure can be expressed as follows:

$$E_{tot} = -K_1^{eff} \cos^2 \theta_M - K_2 \cos^4 \theta_M - M_S \boldsymbol{m} \cdot (\boldsymbol{H}_{ext} + \Delta \boldsymbol{H}), \tag{1}$$

$$\boldsymbol{m} \equiv (\sin \theta_M \cos \varphi_M, \sin \theta_M \sin \varphi_M, \cos \theta_M), \tag{2}$$

$$\boldsymbol{H}_{ext} \equiv H_{ext} (\sin \theta_H \cos \varphi_H, \sin \theta_H \sin \varphi_H, \cos \theta_H), \tag{3}$$

$$\Delta \boldsymbol{H} \equiv \Delta H_{DL} \boldsymbol{m} \times \hat{y} - \Delta H_{FL} \hat{y}. \tag{4}$$

Here, $K_1^{\text{eff}}$ is the effective first-order PMA energy density that takes into account of the demagnetizing term: i.e., $K_1^{\text{eff}} = K_1 - N_d M_S^2/2$ ($K_1$, $N_d$, and $M_S$ are the first-order PMA energy density, demagnetizing factor, and saturation magnetization, respectively).[30] $K_2$ is the second-order PMA energy density.[30] $\boldsymbol{m}$ is the unit vector of magnetization. The effective magnetic field ($\Delta \boldsymbol{H}$) induced by the maximum value of the in-plane AC current ($I_0$) is composed of the damping-like effective field ($\Delta H_{DL}$) and the field-like effective field ($\Delta H_{FL}$). $\theta_H$ and $\varphi_H$ are the polar and azimuthal angles of $\boldsymbol{H}_{ext}$. Given that the in-plane anisotropy is negligibly small over the PMA field, $\varphi_M{}^\circ$ is assumed to be identical to $\varphi_H$. The values of $\Delta \theta_M$ and $\Delta \varphi_M$ can be analytically expressed as follows [refer to Supplementary Eqs. (S1)−(S13) for a detailed derivation]:

$$\Delta \theta_M = -\frac{\Delta H_{DL} \cos \varphi_H + \Delta H_{FL} \cos \theta_M^\circ \sin \varphi_H}{H_K^{eff} \cos 2\theta_M^\circ - H_{K,2} \sin \theta_M^\circ \sin 3\theta_M^\circ + H_{ext} \cos(\theta_M^\circ - \theta_H)}, \tag{5}$$

$$\Delta \varphi_M = \frac{\Delta H_{DL} \cos \theta_M^\circ \sin \varphi_H - \Delta H_{FL} \cos \varphi_H}{H_{ext} \sin \theta_H}. \tag{6}$$

Here, $H_K^{\text{eff}}$ and $H_{K1}^{\text{eff}}$ are the effective PMA field and effective first-order PMA field, respectively and $H_{K,2}$ is the second-order PMA field. These parameters are defined as follows: $H_K{}^{eff} \equiv H_{K,1}{}^{eff} + H_{K,2}$; $H_{K,1}{}^{eff} \equiv 2K_1{}^{eff}/M_S$; $H_{K,2} \equiv 4K_2/M_S$. Note that Eqs. (5) and (6) are identical to the analytical expressions derived by Hayashi *et al.* when $H_{K,2} = 0$.[23] If the values of $\Delta\theta_M$ and $\Delta\varphi_M$ are sufficiently small, the components of the $\boldsymbol{m}$ vector can be approximated in the form $m(t) = m^\circ + (2\Delta m)\sin\omega t$:

$$m_x \approx \sin\theta_M^\circ \cos\varphi_H + \left(\Delta\theta_M \cos\theta_M^\circ \cos\varphi_H - \Delta\varphi_M \sin\theta_M^\circ \sin\varphi_H\right)\sin\omega t, \quad (7)$$

$$m_y \approx \sin\theta_M^\circ \sin\varphi_H + \left(\Delta\theta_M \cos\theta_M^\circ \sin\varphi_H + \Delta\varphi_M \sin\theta_M^\circ \cos\varphi_H\right)\sin\omega t, \quad (8)$$

$$m_z \approx \cos\theta_M^\circ - \Delta\theta_M \sin\theta_M^\circ \sin\omega t. \quad (9)$$

Both the anomalous and the planar Hall voltages contribute to the measured Hall voltage, $V_H = I_{AC} R_H = I_{AC} R_{AHE} m_z + I_{AC} R_{PHE} m_x m_y$.[20,31,32] Here, $R_{AHE}$ and $R_{PHE}$ are the anomalous and planar Hall resistances, respectively. Under application of $I_{AC}$, the $m$ values oscillate as given in Eqs. (7)−(9), with the resultant expressions for Hall voltages being as follows:

$$V_H = \left(V_{AHE} m_z + V_{PHE} m_x m_y\right)\sin\omega t = V^{1\omega}\sin\omega t - V^{2\omega}\cos 2\omega t, \quad (10)$$

$$V_x^{1\omega} = V_y^{1\omega} = V_{AHE}\cos\theta_M^\circ, \quad (11)$$

$$V_x^{2\omega} = -\frac{V_{AHE}}{2}\left[\sin\theta_M^\circ \Delta\theta_M - R\sin^2\theta_M^\circ \Delta\varphi_M\right], \quad (12)$$

$$V_y^{2\omega} = -\frac{V_{AHE}}{2}\left[\sin\theta_M^\circ \Delta\theta_M + R\sin^2\theta_M^\circ \Delta\varphi_M\right]. \quad (13)$$

Here, the following relations exist: $V_{AHE} = I_0 R_{AHE}$ and $V_{PHE} = I_0 R_{PHE}$. The subscripts $x$ and $y$ indicate that the harmonic Hall voltages are measured at $\varphi_H =$ 0° and 90°, respectively. The first harmonic Hall voltage ($V^{1\omega}$) contains information on the $\theta_M$° value, whereas the second harmonic Hall voltage ($V^{2\omega}$) contains information on the $\Delta\theta_M$ and $\Delta\varphi_M$ values. The conventional analytical solution considers only that case in which the magnetization direction has deviated just slightly from the $z$-axis ($\theta_M° \approx 0°$). In this case, the $z$-component of $H_\text{ext}$ is negligibly small over the PMA field along the same direction ($H_\text{ext}\cos\theta_H \ll H_\text{K}^{eff}\cos\theta_M°$), and therefore, the assumption of $\sin\theta_M° = H_\text{ext}/H_\text{K}^{eff}$ made in the conventional solution is reasonable.[8,23,33] Note that the contribution due to $H_{K,2}$ is also negligible at $\theta_M° \approx 0°$ [Eq. (5)]. Under this assumption, $V^{1\omega}$ and $V^{2\omega}$ can be rewritten as follows:

$$V_x^{1\omega} = V_y^{1\omega} = V_\text{AHE}\sqrt{1-\left(\frac{H_\text{ext}}{H_\text{K}^{eff}}\right)^2},$$
(14)

$$V_x^{2\omega} = \frac{V_\text{AHE}H_\text{ext}}{2\left(H_\text{K}^{eff}\right)^2}\left[\frac{\Delta H_\text{DL}}{1-\left(H_\text{ext}/H_\text{K}^{eff}\right)^2}-R\Delta H_\text{FL}\right],$$
(15)

$$V_y^{2\omega} = \frac{V_\text{AHE}H_\text{ext}}{2\left(H_\text{K}^{eff}\right)^2}\sqrt{1-\left(\frac{H_\text{ext}}{H_\text{K}^{eff}}\right)^2}\left[\frac{\Delta H_\text{FL}}{1-\left(H_\text{ext}/H_\text{K}^{eff}\right)^2}-R\Delta H_\text{DL}\right].$$
(16)

The second harmonic Hall voltages, as given in Eqs. (15) and (16), are composed of two terms containing $\Delta H_\text{DL}$ and $\Delta H_\text{FL}$. When the $R$ ratio is negligibly small, the values of $\Delta H_\text{DL}$ and $\Delta H_\text{FL}$ can be obtained using the following $T$ ratios:

$$T_x \equiv -\frac{2V_x^{2\omega}}{\left(\partial V_x^{1\omega}/\partial H_\text{ext}\right)}\left(\frac{V_x^{1\omega}}{V_\text{AHE}}\right) = A_0\Delta H_\text{DL} - B_0\Delta H_\text{FL},$$
(17)

$$T_y \equiv \frac{2V_y^{2\omega}}{\left(\partial V_y^{1\omega}/\partial H_{\text{ext}}\right)} = B_0 \Delta H_{\text{DL}} - A_0 \Delta H_{\text{FL}}, \tag{18}$$

$$A_0 \equiv 1, \qquad B_0 \equiv R\left(1 - \left(\frac{H_{\text{ext}}}{H_{\text{K}}^{eff}}\right)^2\right). \tag{19}$$

Note that at $R = 0$, the values of $T_x$ and $T_y$ are identical to those of $\Delta H_{\text{DL}}$ and $-\Delta H_{\text{FL}}$, respectively.[8] When the $R$ ratio becomes comparatively large, $T_x$ and $T_y$ should be corrected using Cramer's rule.[23]

$$\begin{pmatrix} \Delta H_{\text{DL}} \\ \Delta H_{\text{FL}} \end{pmatrix} = \frac{1}{B_0^2 - A_0^2} \begin{pmatrix} -A_0 & B_0 \\ -B_0 & A_0 \end{pmatrix} \begin{pmatrix} T_x \\ T_y \end{pmatrix}. \tag{20}$$

In Eq. (20), $\Delta H_{\text{DL}}$ and $\Delta H_{\text{FL}}$ can be calculated if the determinant $B_0^2 - A_0^2$ is not 0. If $B_0^2 - A_0^2$ is 0, it is not possible to obtain the individual values of $\Delta H_{\text{DL}}$ and $\Delta H_{\text{FL}}$; it is rather possible to obtain only the relation $T_x = T_y = \Delta H_{\text{DL}} - \Delta H_{\text{FL}}$ [Eqs. (17)−(19)].

### B. Analytical solutions for refined approach

The assumption of $H_{\text{ext}} \cos\theta_H \ll H_{\text{K}}^{eff} \cos\theta_M{}^\circ$ is no longer valid at high $H_{\text{ext}}$ values. In this case, the $\theta_H$ value is not negligible, and it is then necessary to substitute $\Delta\theta_M$ and $\Delta\varphi_M$ [Eqs. (5) and (6)] into Eqs. (12) and (13) to obtain the expression for $V^{2\omega}$:

$$V_x^{2\omega} = \frac{V_{\text{AHE}}}{2}\left[A_1 \Delta H_{\text{DL}} - B_1 \Delta H_{\text{FL}}\right], \tag{21}$$

$$V_y^{2\omega} = -\frac{V_{\text{AHE}}\cos\theta_M^o}{2}\left[ B_1\Delta H_{\text{DL}} - A_1\Delta H_{\text{FL}} \right], \tag{22}$$

$$A_1 \equiv \frac{\sin\theta_M^o}{H_K^{eff}\cos 2\theta_M^o - H_{K,2}\sin\theta_M^o \sin 3\theta_M^o + H_{\text{ext}}\cos\left(\theta_M^o - \theta_H\right)}, \tag{23}$$

$$B_1 \equiv \frac{R\sin^2\theta_M^o}{H_{\text{ext}}\sin\theta_H}. \tag{24}$$

Considering that $V_x^{1\omega} = V_y^{1\omega} = V_{\text{AHE}}\cos\theta_M^o$ [Eq. (11)], the $G$ ratios, corresponding to the $T$ ratios used in the conventional approach, can be defined as follows:

$$G_x \equiv \frac{2V_x^{2\omega}}{V_{\text{AHE}}} = A_1\Delta H_{\text{DL}} - B_1\Delta H_{\text{FL}}, \tag{25}$$

$$G_y \equiv -\frac{2V_y^{2\omega}}{V_y^{1\omega}} = B_1\Delta H_{\text{DL}} - A_1\Delta H_{\text{FL}}. \tag{26}$$

Similarly to the conventional analytical equations, the refined equations also need to be solved using Cramer's rule [as given in Eq. (20)].

$$\begin{pmatrix} \Delta H_{\text{DL}} \\ \Delta H_{\text{FL}} \end{pmatrix} = \frac{1}{B_1^{\ 2} - A_1^{\ 2}} \begin{pmatrix} -A_1 & B_1 \\ -B_1 & A_1 \end{pmatrix} \begin{pmatrix} G_x \\ G_y \end{pmatrix}. \tag{27}$$

## RESULTS

### A. Conventional analysis

The conventional analytical method is used to analyze the results of the macrospin simulation. Figures 1(a) and (b) show the results for $V^{1\omega}$ as a function of $H_{ext}$ in two different systems: (a) $H_{K,1}{}^{eff} = 5$ kOe and $H_{K,2} = 0$; (b) $H_{K,1}{}^{eff} = 5$ kOe and $H_{K,2} = -1$ kOe. Two sets of results are shown in Figs. 1(a) and (b): one is obtained from the macrospin simulation (squares) and the other from Eq. (14), which is based on the conventional assumption of $\sin\theta_M{}^\circ = H_{ext}/H_K{}^{eff}$ (dashed lines). For the macrospin simulation, the following parameters are used: $\Delta H_{DL} = -50$ Oe, $\Delta H_{FL} = -100$ Oe, $\theta_H = 86°$, and $V_{AHE} = 1$ mV. Refer to Supplementary Fig. S1 for a detailed description on the macrospin simulation. The agreement between the results obtained from the macrospin simulation and those obtained from Eq. (14) based on the conventional analytical method, which is the main focus of this study, is good only in the low-$H_{ext}$ range. In the high-$H_{ext}$ range, the deviation is indeed very large, indicating the limited validity of the conventional solutions.

Figures 1(c) and (d) show the analytical results for $B_0{}^2 - A_0{}^2$ calculated from Eq. (19) as a function of $H_{ext}$ at two different $R$ values of 0.3 (red lines) and 1.75 (blue lines) ($R = V_{PHE}/V_{AHE}$). The results in Fig. 1(c) are for the system with $H_{K,1}{}^{eff} = 5$ kOe and $H_{K,2} = 0$ and those in Fig. 1(d) are for the system with $H_{K,1}{}^{eff} = 5$ kOe and $H_{K,2} = -1$ kOe. The $H_K{}^{eff}$ values for both the systems are also indicated in the figures. The detailed equation for $B_0{}^2 - A_0{}^2$ is rewritten just for clarity:

$$B_0{}^2 - A_0{}^2 = R^2 \left( 1 - \left( \frac{H_{ext}}{H_K^{eff}} \right)^2 \right)^2 - 1.$$

(28)

Recalling that $H_{ext}/H_K{}^{eff}$ is approximated with $\sin\theta_M{}^\circ$, we can say that the results for $B_0{}^2 - A_0{}^2$ at $H_{ext} > H_K{}^{eff}$ have no physical meaning. According to Eq. (28), the $B_0{}^2 - A_0{}^2$ value decreases from $R^2 - 1$ to $-1$ as the $H_{ext}$ value increases from 0 to $H_K{}^{eff}$. More specifically, the determinant $B_0{}^2 - A_0{}^2$ always has a negative value at $R < 1$. At $R \geq 1$, however, the determinant can have both positive and negative values over the $H_{ext}$ range of $0$–$H_K{}^{eff}$, indicating the occurrence of a crossover $(B_0{}^2 - A_0{}^2 = 0)$ at a certain $H_{ext}$ value. This feature is visible clearly in the results shown in Figs. 1(c) and (d). In both the systems, i.e., with $H_{K,2} = 0$ and $-1$ kOe, the $B_0{}^2 - A_0{}^2$ value is always negative at $R = 0.3$, but at $R = 1.75$, it initially has a positive value, after which it passes through 0 and then finally, it becomes a negative value. The crossovers occur at 3.3 and 2.6 kOe for the systems with $H_{K,2} = 0$ and $-1$ kOe, respectively. Recalling that $T_x = T_y = \Delta H_{DL} - \Delta H_{FL}$ when the determinant is 0, an $H_{ext}$ value should exist at which $T_x = T_y$ when $R > 1$.

Figures 2(a)$-$(f) show the results for $V_x{}^{2\omega}$ and $V_y{}^{2\omega}$ [(a) and (b)] and $T_x$ and $T_y$ [(c) and (d)] as functions of $H_{ext}$ and those for $\Delta H_{DL}$ and $\Delta H_{FL}$ [(e) and (f)] as functions of $\theta_M{}^\circ$ for two different systems, i.e., with $H_{K,2} = 0$ (solid lines) and $-1$ kOe (dashed lines). The left [Figs. 2(a), (c), and (e)] and right [Figs. 2(b), (d), and (f)] panels show the results for $R = 0.3$ and $R = 1.75$, respectively. The results for $V^{2\omega}$ were obtained by the macrospin simulation and those for $T_x$ and $T_y$ [Eqs. (17) and (18)] and $\Delta H_{DL}$ and $\Delta H_{FL}$ [Eq. (20)] were calculated analytically using the simulation results. It is seen from Figs. 2(a) and (b) that the sign of $V_x{}^{2\omega}$ at a small $R$ value of 0.3 is negative, but it is positive at a large $R$ value of 1.75. This is because, between the two major contributions of $V_{AHE}$ and $V_{PHE}$ to $V^{2\omega}$, the sign of the former is negative, but that of the latter is positive. Indeed, Eqs. (15) and (16) predict this behavior (the opposite signs of $V_{AHE}$ and $V_{PHE}$) and furthermore, explain that the $V_y{}^{2\omega}$ value at $R = 1.75$ is higher than that at $R = 0.3$.

The results for $V_x^{2\omega}$ and $V_y^{2\omega}$ and for their variation with $R$ have a critical effect on $T_x$ and $T_y$. At $R = 0.3$, the signs of $T_x$ and $T_y$ are opposite because the signs of $V_x^{2\omega}$ and $V_y^{2\omega}$ are the same, indicating that there are no $H_{ext}$ values at which $T_x = T_y$ in both the systems, i.e., with $H_{K,2} = 0$ and $-1$ kOe [Fig. 2(c)]. These results are consistent with those for $B_0^2 - A_0^2$ [Figs. 1(c) and (d)]. It should be remembered that the values of $T_x$ and $T_y$ are the same at a specific $H_{ext}$ value at which $B_0^2 - A_0^2 = 0$ [Eq. (19)]. At $R = 1.75$, the signs of $T_x$ and $T_y$ are the same because the signs of $V_x^{2\omega}$ and $V_y^{2\omega}$ are opposite [Fig. 2(d)]. $H_{ext}$ values at which $T_x = T_y$ exist in both the systems, i.e., with $H_{K,2} = 0$ and $-1$ kOe. The positions, however, are quite different from those at which $B_0^2 - A_0^2 = 0$. The $H_{ext}$ values in the former case are 3.6 and 3.8 kOe for the systems with $H_{K,2} = 0$ and $-1$ kOe, respectively, whereas those in the latter case are 3.3 and 2.6 kOe, respectively. These deviations occur because the determinant poorly reflects the behavior of the first harmonic.

The inappropriate determinant, i.e., $B_0^2 - A_0^2$, causes large errors in the SO effective fields, as shown in Figs. 2(e) and (f). Recalling that the input SO effective fields are $\Delta H_{DL} = -50$ Oe and $\Delta H_{FL} = -100$ Oe, we can consider the results at $R = 0.3$ [Fig. 2(e)] to be quite reliable in the $\theta_M°$ range from $0°$ to the angles corresponding to $H_{ext} = H_K^{eff}$. These angles are $61°$ and $52°$ when $H_{K,2} = 0$ and $H_{K,2} = -1$ kOe, respectively. Beyond these two angles, which are indicated by vertical and horizontal lines, respectively, in Figs. 2(e) and (f), the output SO effective fields start to deviate from the input values. The indicated regions end not at $90°$ but at $\sim 82°$, since the $\boldsymbol{m}$ vector is not fully aligned along the $x$-axis or the $y$-axis even under $H_{ext} = 10$ kOe [Figs. 1(a) and (b)]. The output SO effective fields show a divergence, which is physically meaningless, at $\theta_M° = $ ~81° ($H_{K,2} =$ 0 kOe) and ~75° ($H_{K,2} = -1$ kOe). The deviations are very large at $R = 1.75$ [Fig. 2(f)]. For the system with $H_{K,2} = 0$ kOe, the divergence occurs even at ~37°, which does not lie in the region of physical insignificance (indicated by the vertical lines). A similar behavior is observed for the system with $H_{K,2} = -1$ kOe, where the divergence occurs at ~32°. These divergences are attributed to the mislocation of the $H_{ext}$ value at which $B_0{}^2 - A_0{}^2 = 0$ [Figs. 1(c) and (d)]. The occurrence of the additional divergences significantly limits the reliability of the conventional analytical method for both the systems, i.e., with $H_{K,2} = 0$ and $-1$ kOe, as can be seen clearly from Fig. 2(f).

## B. Refined analysis

The main reason behind the unreliable results obtained from the conventional analysis is the determinant, which poorly describes the behavior of the first harmonic Hall voltage. For an accurate evaluation of the determinant, the refined analysis is begun with determination of the relation between $\theta_M°$ and $\boldsymbol{H}_{ext}$, which can be obtained using the equation $\theta_M° = \cos^{-1}(V^{1\omega}/V_{AHE})$ or the total energy equation[34] [Eq. (1)]. The results for $V^{1\omega}$ as a function of $H_{ext}$ [Figs. 1(a) and (b)] can be used to obtain the relation. Figures 3(a) and (b) show the results for the determinant $B_1{}^2 - A_1{}^2$ as a function of $H_{ext}$ obtained from the refined analysis [Eqs. (23) and (24)] for the systems with $H_{K,2} = 0$ and $-1$ kOe, respectively. Solid lines in Figs. 3(a) and (b) indicate the results for $B_1{}^2 - A_1{}^2$ calculated by using $H_{K,2} = 0$ and $-1$ kOe from the macrospin simulation and the refined analysis, respectively. Use of the relation between $\theta_M°$ and $\boldsymbol{H}_{ext}$ in the refined analysis leads to the behavior of $V^{1\omega}$ being duly reflected in the determinant. At $R = 1.75$, the $H_{ext}$ values at which $B_1^2 - A_1^2 = 0$ are 3.6 and 3.8 kOe for the systems with $H_{K,2} = 0$ and $-1$ kOe, respectively; these $H_{ext}$ values are identical to those obtained at $T_x = T_y$ [Fig. 2(d)]. To apply the behavior of $V^{1\omega}$ to the determinant, the determinant $B_1^2 - A_1^2$ needs to be calculated with a precise value of $H_{K,2}$, which was used in the macrospin simulation. In order to demonstrate the importance of the inclusion of $H_{K,2}$, the determinants were also calculated by ignoring $H_{K,2}$ (even though the actual $H_{K,2}$ value of the system is $-1$ kOe); these results are also shown in Fig. 3(b) (dotted lines). Indeed, the difference is very large between the two cases of $R = 0.3$ and 1.75, indicating that $H_{K,2}$ should be taken into account in the analysis. At $R = 1.75$, for example, the $H_{ext}$ value at which the determinant is 0 is mislocated from 3.8 to 3.2 kOe when $H_{K,2}$ is ignored; furthermore, a new location showing a 0 value of the determinant emerges at $H_{ext} = 9.0$ kOe.

Figures 4(a) and (b) show the results for $G_x$ and $G_y$, which correspond to $T_x$ and $T_y$ in the conventional analysis, at $R = 0.3$ and 1.75, respectively. The results are shown for the systems with $H_{K,2} = 0$ (solid lines) and $-1$ kOe (dashed lines). Note that the $H_{ext}$ values at which $G_x = G_y$ are 3.6 and 3.8 kOe for the systems with $H_{K,2} = 0$ and $-1$ kOe, respectively. These $H_{ext}$ values are identical to those at which the determinant $B_1^2 - A_1^2 = 0$ [Figs. 3(a) and (b)]. This is in contrast to the case of the conventional analysis, where the $H_{ext}$ value at which the determinant is 0 differs substantially from that at which $T_x = T_y$ [Figs. 1(c) and (d) and Fig. 2(d)]. Armed with the new set of results for the determinant and the $G$ ratios, it is a straightforward task for us to calculate the SO effective fields;

these results as a function of $\theta_M°$ are shown in Figs. 4(c) and (d) for $R = 0.3$ and 1.75, respectively. Two sets of results are shown: one each for the systems with $H_{K,2} = 0$ (solid lines) and $-1$ kOe (dashed lines). It is seen from Figs. 4(c) and (d) that in both these systems, the calculated values of $\Delta H_{DL}$ and $\Delta H_{FL}$ are in excellent agreement with the input values used for the macrospin simulation (over the entire $\theta_M°$ range of $0°$ to $\sim82°$); this demonstrates the reliability of the refined analysis. Specifically, at $R = 0.3$, the agreement is perfect between the two systems to such an extent that the solid lines for the $H_{K,2} = 0$ kOe system overlap completely with the dashed lines for the $H_{K,2} = -1$ kOe system over the entire $\theta_M°$ range. A similar behavior is observed at $R = 1.75$, the only difference being that small peaks are observed at $\sim43°$ and $\sim50°$ for the systems with $H_{K,2} = 0$ and $-1$ kOe, respectively, at which $B_1^2 - A_1^2 = 0$.

In systems having both $H_{K,1}{}^{eff}$ and $H_{K,2}$, the determinant $B_1^2 - A_1^2$ differs significantly if $H_{K,2}$ is ignored [Fig. 3(b)]. A similar difference is then expected in the calculated values of $\Delta H_{DL}$ and $\Delta H_{FL}$ [using Eq. (27)], which are also shown in Figs. 4(c) and (d) (dotted lines). At $R = 0.3$, the absolute values of $\Delta H_{DL}$ and $\Delta H_{FL}$ are underestimated in the $\theta_M°$ range of $0°$ to $60°$ and overestimated in the range of $60°$ to $\sim82°$. This can be understood from the $H_{K,2}$ term, which is proportional to $\sin\theta_M°\sin3\theta_M°$ [Eq. (23)]. At $R = 1.75$, the differences are rather huge, with two divergences: at $\sim39°$ and $\sim80°$. This is mainly due to the mislocated $H_{ext}$ fields of 3.2 and 9.0 kOe at which $B_1^2 - A_1^2 = 0$ [Fig. 3(b)]. These results clearly demonstrate that in systems having both $H_{K,1}{}^{eff}$ and $H_{K,2}$, $H_{K,2}$ should not be neglected in the analysis of the harmonic measurement results.

## C. Comparison of conventional and refined analyses over wide $R$ range

Two typical $R$ ratios of 0.3 and 1.75 have been considered thus far. In order to test the refined analytical method over a wide $R$ range, a more systematic study was conducted by varying the $R$ ratio from 0 to 2 in steps of 0.05 for the system with $H_{K,2} = -1$ kOe. Figures 5(a) and (b) display contour plots showing the deviation (in %) from the input values of $\Delta H_{DL}$ (left panels) and $\Delta H_{FL}$ (right panels) as a function of $\theta_M{}^\circ$ and $R$. The results calculated using the conventional analytical method are shown in Fig. 5(a), whereas those using the refined method are shown in Fig. 5(b). In the case of the conventional solutions, the $\theta_M{}^\circ$ range in which $H_{ext} > H_K{}^{eff}$ has no physical significance is indicated in Fig. 5(a) as inclined lines. Furthermore, in Figs. 5(a) and (b), the solid lines indicate a deviation of 0.8% and the white regions indicate a deviation of 4% or larger. It is seen from Fig. 5(a) that the conventional solutions are valid over very limited ranges of $\theta_M{}^\circ$ and $R$. For example, the $R$ range in which the deviations are less than 4% is 0.06–0.12 for $\Delta H_{DL}$ and 0.21–0.46 for $\Delta H_{FL}$ in the $\theta_M{}^\circ$ range of 0°–52°. Furthermore, at $R$ values higher than 1.1, the validity range is even more limited for both $\Delta H_{DL}$ and $\Delta H_{FL}$; specifically, the $\theta_M{}^\circ$ values at which the deviations are less than 4% are 4.5° at $R = 1.1$ and 7.9° at $R = 2.0$ for $\Delta H_{DL}$, and they are 4.5° at $R = 1.1$ and 9.4° at $R = 2.0$ for $\Delta H_{FL}$. In the intermediate $R$ range of 0.9–1.1, the deviations are always larger than 4%. The accuracy of the calculated results improves significantly with the use of the refined method, as can be seen from Fig. 5(b). With the $z$-component of $\boldsymbol{H}_{ext}$ taken into account in the refined analysis, there is no region of physical insignificance. Furthermore, the predictions made in the refined analysis are highly accurate. At $R < 0.85$, the deviations are less than 0.4% over the entire $\theta_M°$ range of 0°–82° for both $\Delta H_{DL}$ and $\Delta H_{FL}$. Even at $R > 0.85$, the deviations are less than 0.8% for both $\Delta H_{DL}$ and $\Delta H_{FL}$ over the entire region, except in the regions marked by solid lines, where the deviations are rather large owing to the existence of divergences (zero determinants).

## D. Analysis of experimental results

A further test of the refined method was made by analyzing the experimental results of harmonic Hall voltage measurements for a stack with the structure: Si substrate (wet-oxidized)/ Ta (5 nm)/Pt (5 nm)/ Co (0.6 nm)/MgO (2 nm)/Ta (2 nm). Refer to Supplementary Figs. (S2) and (S3) for Hall bar dimensions. The results were obtained at three different $I_0$ of 1.0, 1.5, and 2.0 mA. The magnetization direction was controlled by $H_{ext}$, which was swept from +90 to −90 kOe with two different directions of $\theta_H = 85°$ and $\varphi_H = 0°$, and $\theta_H = 85°$ and $\varphi_H = 90°$. The values of $H_{K,1}{}^{eff}$ and $H_{K,2}$, which were extracted using the Generalized Sucksmith–Thompson method[35], were 33.1 kOe and −8.1 kOe, respectively. The $R$ ratio of the sample was measured to be 0.423. Figures 6(a)−(d) show the results for $\Delta H_{DL}$ [(a) and (c)] and $\Delta H_{FL}$ [(b) and (d)] as functions of $\theta_M°$ at three different $I_0$ values of 1.0 mA (black squares), 1.5 mA (red circles), and 2.0 mA (blue triangles). Both the conventional [(a) and (b)] and refined [(c) and (d)] analytical methods were used to analyze the experimental results. The results obtained using the conventional method show incorrect divergences at $\theta_M° = {\sim}60°$, but those estimated using the refined method do not show this behavior over the entire $\theta_M°$ range. Note that both $\Delta H_{DL}$ and $\Delta H_{FL}$ depend on $\theta_M{}^\circ$. The values of $\Delta H_{DL}$ and $\Delta H_{FL}$ should be proportional to $I_0$, with zero values at $I_0 = 0$.[15] The expectation is met only for the results extracted using the refined method, as can be seen from Figs. 6(e) and (f), where the results for $\Delta H_{DL}$ (red circles) and $\Delta H_{FL}$ (black squares) obtained at a fixed $\theta_M{}^\circ$ value of $55^\circ$ are shown as functions of $I_0$, respectively. A large deviation from the linearity is particularly noted for the $\Delta H_{FL}$ results calculated using the conventional method. Refer to Supplementary Figs. (S4)–(S6) for detailed results.

## DISCUSSION

The test of the conventional analytical method, which involves analysis of the macrospin simulation results, clearly indicates that its validity range is very limited in terms of $\theta_M{}^\circ$ and $R$; this is due mainly to the singularities involved in Cramer's rule at incorrect $\theta_M{}^\circ$ values. This problem is overcome by the refined analytical method proposed in this study with detailed analytical equations, in which both the $z$-component of $\boldsymbol{H}_{ext}$ and the second-order PMA are taken into account. The SO effective fields extracted using the refined analytical method are in excellent agreement with the input SO effective fields used for the macrospin simulation over the entire $\theta_M{}^\circ$ range and over a wide $R$ range of 0 to 2. Specifically, at $R < 0.85$, deviations from the input SO effective fields are less than 0.4% over the entire $\theta_M{}^\circ$ range of $0^\circ$–$82^\circ$ for both $\Delta H_{DL}$ and $\Delta H_{FL}$. Even at $R > 0.85$, the deviations are less than 0.8% for both $\Delta H_{DL}$ and $\Delta H_{FL}$ over the entire region, except in some limited regions showing singularities. The accuracy of the refined method is confirmed again from an additional comparative study that involves analyzing the experimental results of harmonic Hall voltage measurements for a Pt/Co/MgO structure. An accurate analysis of the harmonic Hall voltage measurement results by the refined analytical method over very wide ranges of $\theta_M{}^\circ$ and $R$ will greatly contribute to the identification of a dominant mechanism of the SOT and the development of highly efficient SOT devices.

## ACKNOWLEDGEMENT

This research was supported by the Creative Materials Discovery Program through the National Research Foundation of Korea (No. 2015M3D1A1070465).

## FIGURE CAPTIONS

FIG. 1. Results for $V^{1\omega}$ [(a) and (b)] and $B_0{}^2 - A_0{}^2$ [(c) and (d)] as functions of $H_{\text{ext}}$ for systems with $H_{\text{K},1}{}^{eff} = 5$ kOe and $H_{\text{K},2} = 0$ kOe [left panels (a) and (c)] and $H_{\text{K},1}{}^{eff} = 5$ kOe and $H_{\text{K},2} = -1$ kOe [right panels (b) and (d)]. For the macrospin simulation, the following parameters are used: $\Delta H_{\text{DL}} = -50$ Oe, $\Delta H_{\text{FL}} = -100$ Oe, $\theta_H = 86°$, and $V_{\text{AHE}} = 1$ mV. In (a) and (b), two sets of results are shown: one from the macrospin simulation (squares) and the other from Eq. (14), which is based on the conventional analytical method (dashed lines). In (c) and (d), two sets of results, both of which are obtained by the conventional analytical method, are shown for $R = 0.3$ (red lines) and 1.75 (blue lines).

FIG. 2. Results for $V_x{}^{2\omega}$ and $V_y{}^{2\omega}$ [(a) and (b)] and $T_x$ and $T_y$ [(c) and (d)] as functions of $H_{\text{ext}}$ and those for $\Delta H_{\text{DL}}$ and $\Delta H_{\text{FL}}$ [(e) and (f)] as functions of $\theta_M°$. The results for $V^{2\omega}$ are obtained by a macrospin simulation and those for $T_x$, $T_y$, $\Delta H_{\text{DL}}$, and $\Delta H_{\text{FL}}$ are obtained by analysis of the simulation results using the conventional analytical method. The results are for the systems with $H_{\text{K},2} = 0$ kOe (solid lines) and $H_{\text{K},2} = -1$ kOe (dashed lines). The left panels [(a), (c), and (e)] show the results for $R = 0.3$, whereas the right panels [(b), (d), and (f)] show those for $R = 1.75$. In (e) and (f), the regions filled with vertical and horizontal lines indicate the $\theta_M°$ range of no physical significance for the systems with $H_{\text{K},2} = 0$ and $-1$ kOe, respectively.

FIG. 3. Refined analytical results for $B_1{}^2 - A_1{}^2$ as a function of $H_{\text{ext}}$ for systems with (a) $H_{\text{K},2} = 0$ kOe and (b) $H_{\text{K},2} = -1$ kOe. Two sets of results, for $R = 0.3$ (red

lines) and $R = 1.75$ (blue lines), are shown. In (b), the results calculated by ignoring $H_{K,2}$ (even though it does exist) are also shown (dotted lines).

FIG. 4. Refined analytical results for $G_x$ and $G_y$ as functions of $H_{ext}$ [(a) and (b)] and those for $\Delta H_{DL}$ and $\Delta H_{FL}$ as functions of $\theta_M{}^\circ$ [(c) and (d)] for systems with $H_{K,2} = 0$ (solid lines) and $H_{K,2} = -1$ kOe (dashed lines). The left [(a) and (c)] and right [(b) and (d)] panels show the results for $R = 0.3$ and $1.75$, respectively. In (c) and (d), the results calculated by ignoring $H_{K,2}$ (even though it does exist) are also shown (dotted lines).

FIG. 5. Contour plots showing deviation (in %) from input values of $\Delta H_{DL}$ (left panel) and $\Delta H_{FL}$ (right panel) as a function of $\theta_M{}^\circ$ and $R$. The results obtained from the conventional analytical method are shown in (a), and those obtained from the refined analytical method are shown in (b). All the results are for the system with $H_{K,2} = -1$ kOe. In (a), the regions filled with inclined lines indicate the $\theta_M{}^\circ$ range of no physical significance.

FIG. 6. Results for $\Delta H_{DL}$ [(a) and (c)] and $\Delta H_{FL}$ [(b) and (d)] as functions of $\theta_M{}^\circ$ at three different $I_0$ values of 1.0 mA (black squares), 1.5 mA (red circles), and 2.0 mA (blue triangles), which are calculated using the conventional [(a) and (b)] and refined [(c) and (d)] analytical methods. Results for $\Delta H_{DL}$ [(e)] and $\Delta H_{FL}$ [(f)] obtained at a $\theta_M{}^\circ$ value of $55°$ as functions of $I_0$ calculated using the conventional (black squares) and refined (red circles) analytical methods. Lines in (e) and (f) are the least squares fit to the refined results.

Graphical Abstract: Schematics of harmonic Hall voltage measurement under external magnetic fields along the $x$-axis (upper) and $y$-axis (lower) with slightly tilted to the $z$-axis. Contour plots showing deviation (in %) from input values of the damping-like ($\Delta H_{\text{DL}}$, left panel) and field-like ($\Delta H_{\text{FL}}$, right panel) spin-orbit effective fields as functions of the angle between the $z$-axis and the magnetization ($\theta_M°$) and the ratio of the planar to the anomalous Hall voltage ($R$). The results obtained from the conventional analytical method are shown in the upper plots, and those obtained from the refined analytical method are shown in the lower plots.

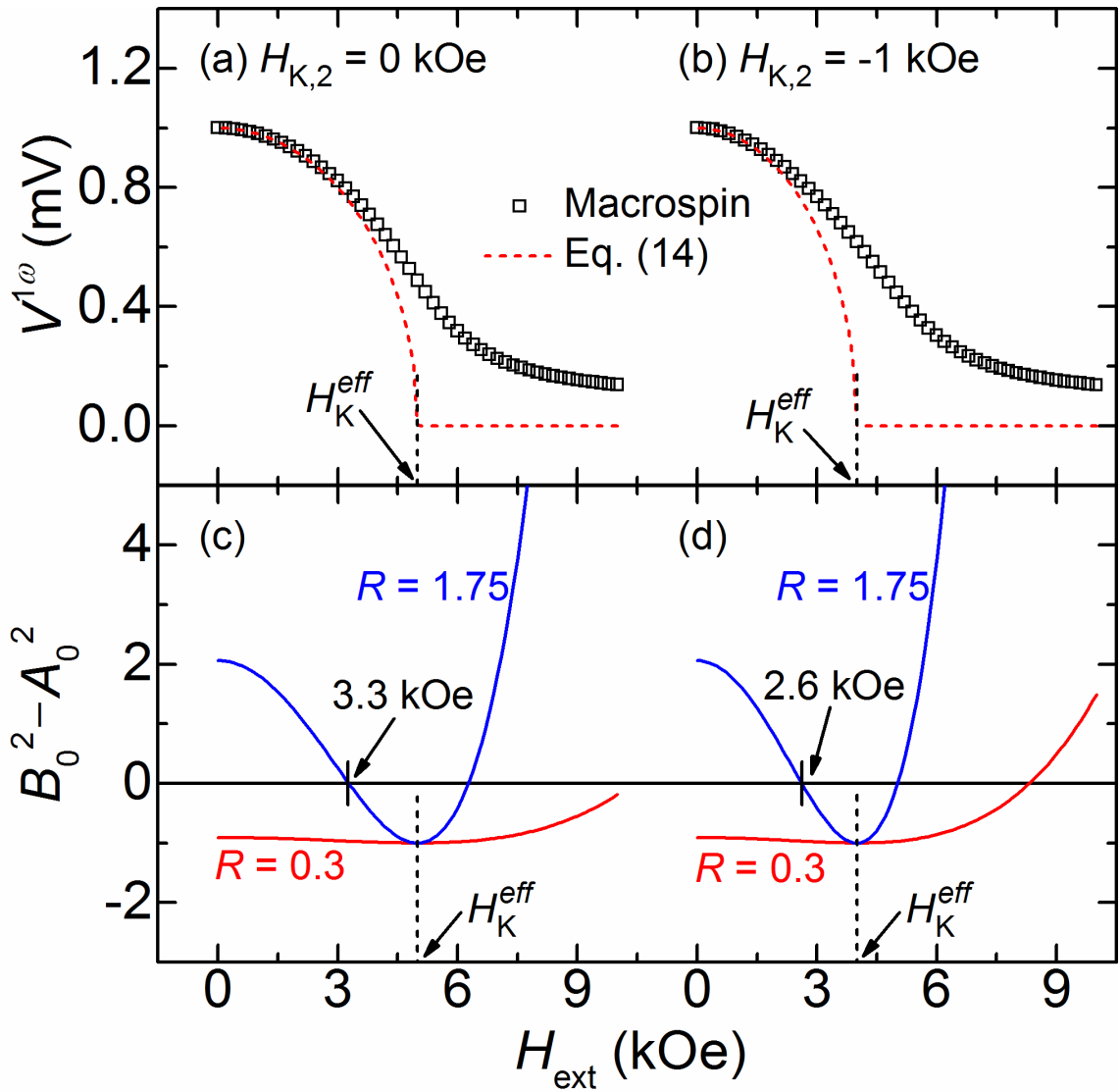

FIG. 1

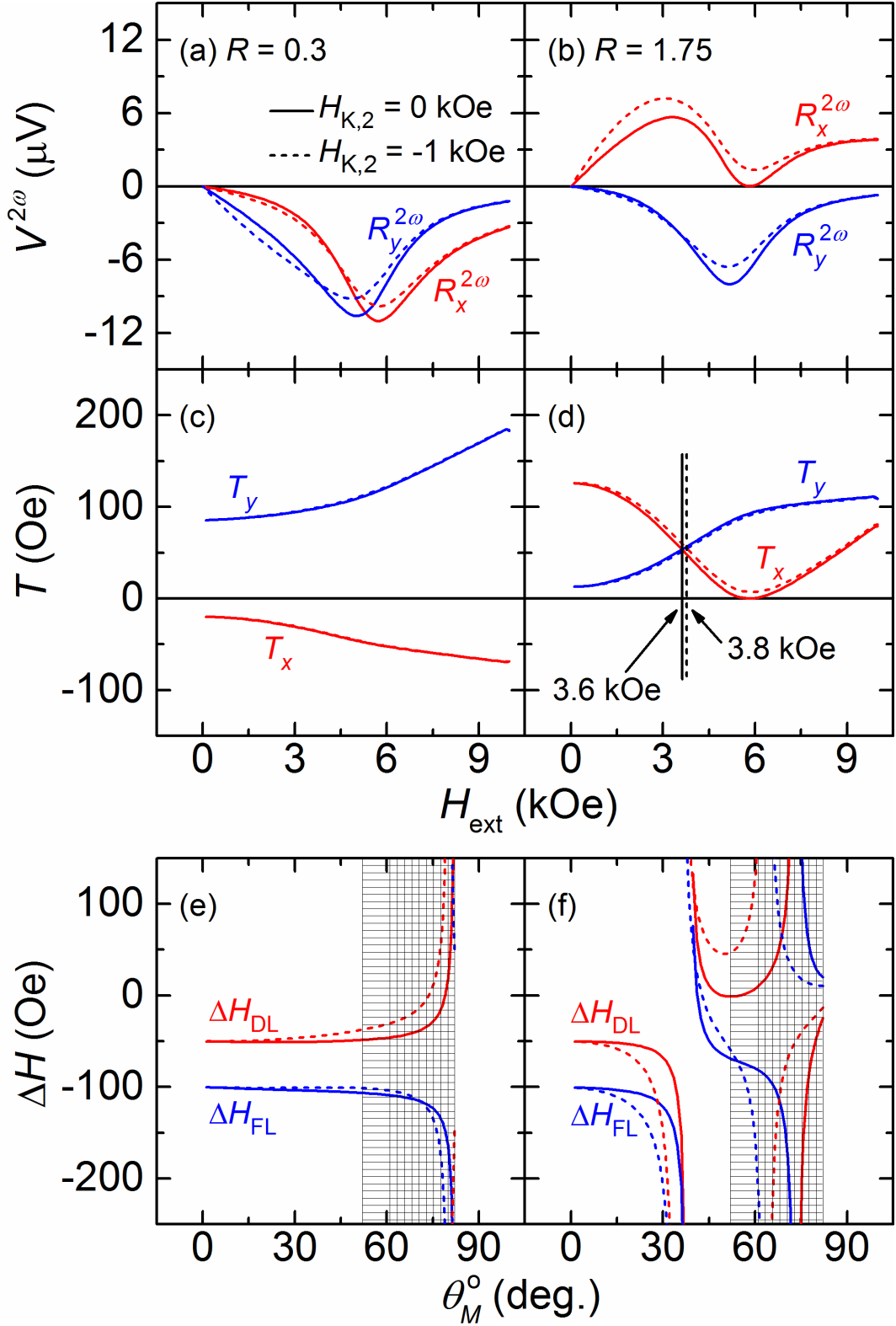

FIG. 2

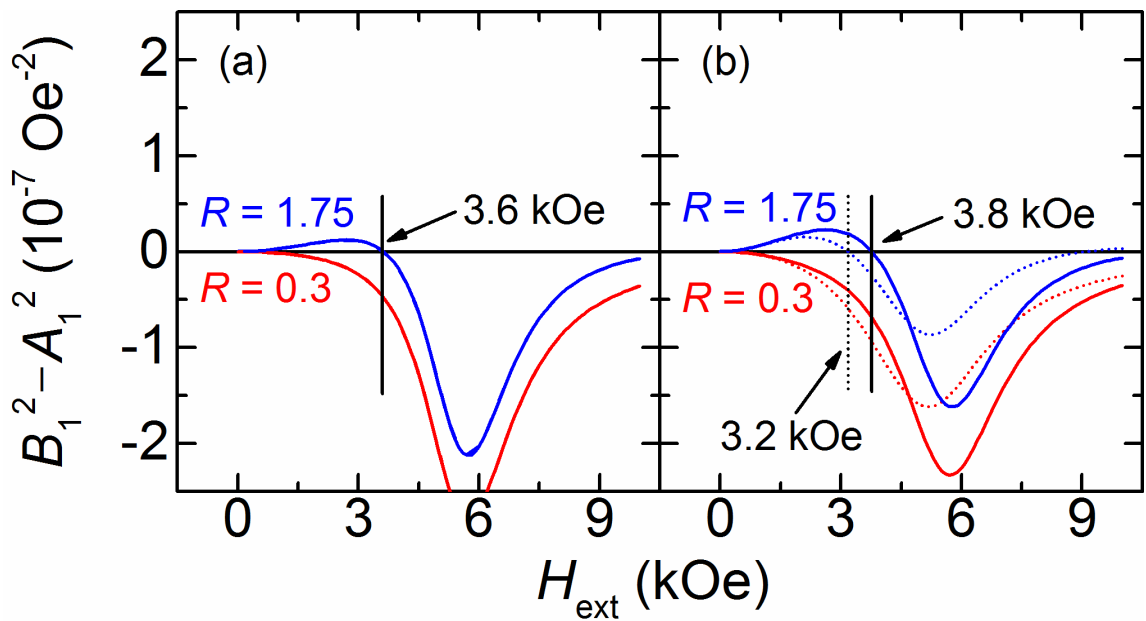

FIG. 3

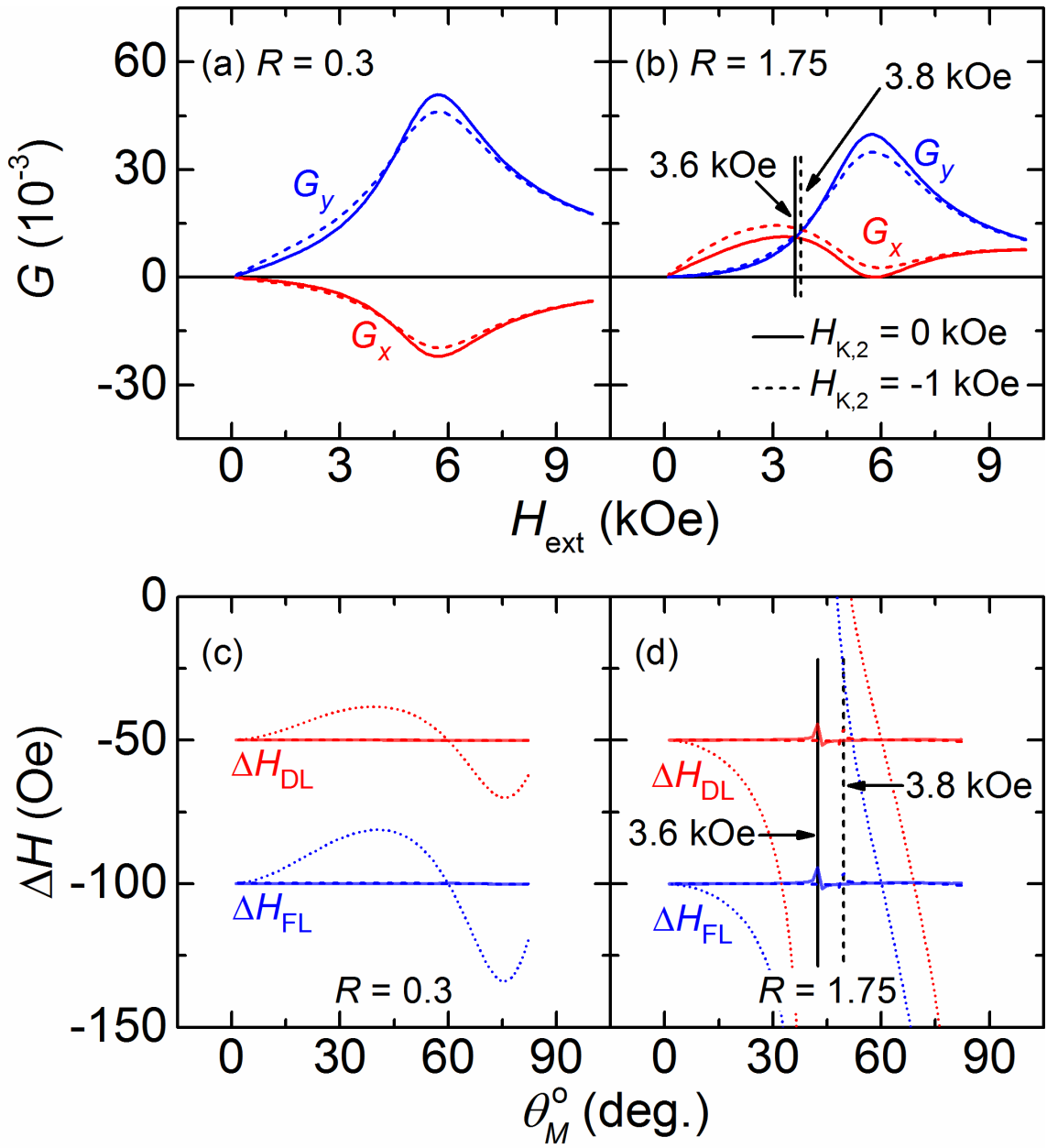

FIG. 4

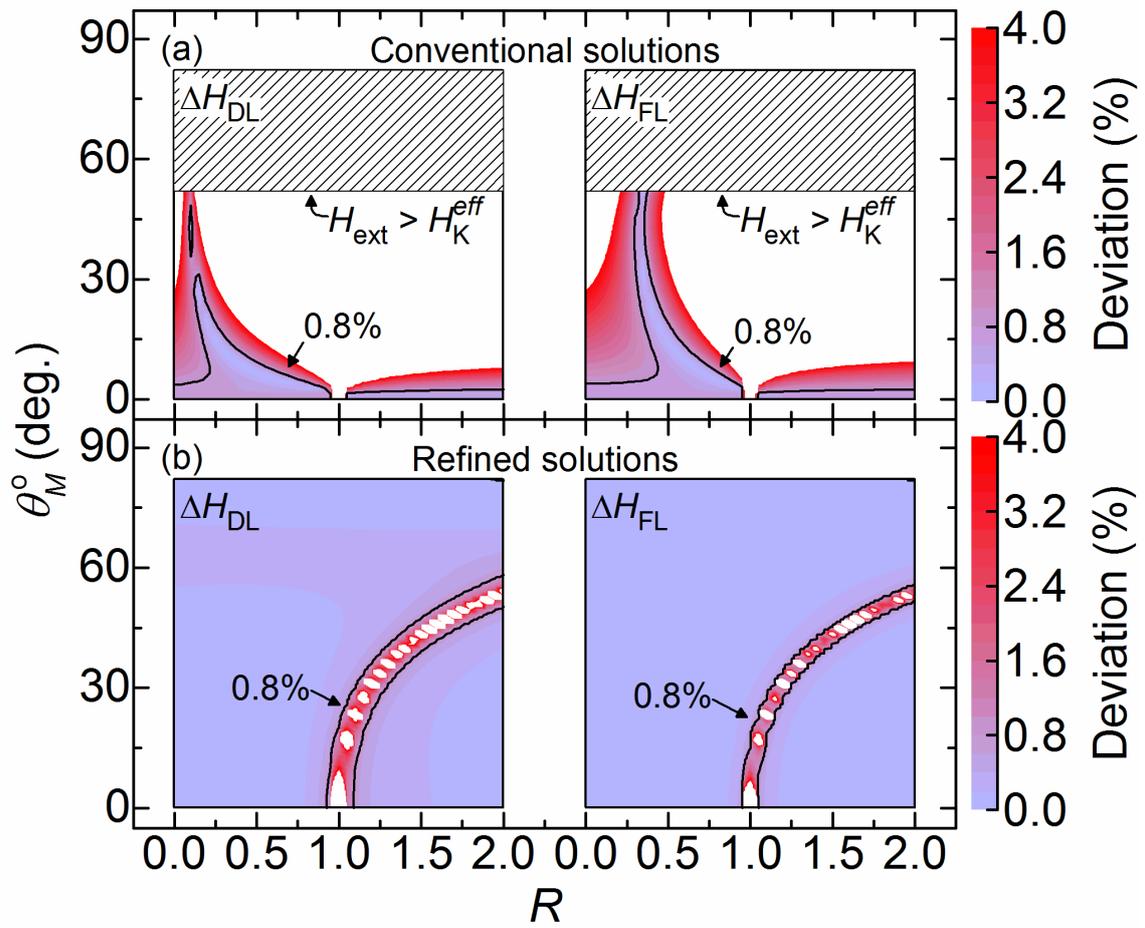

FIG. 5

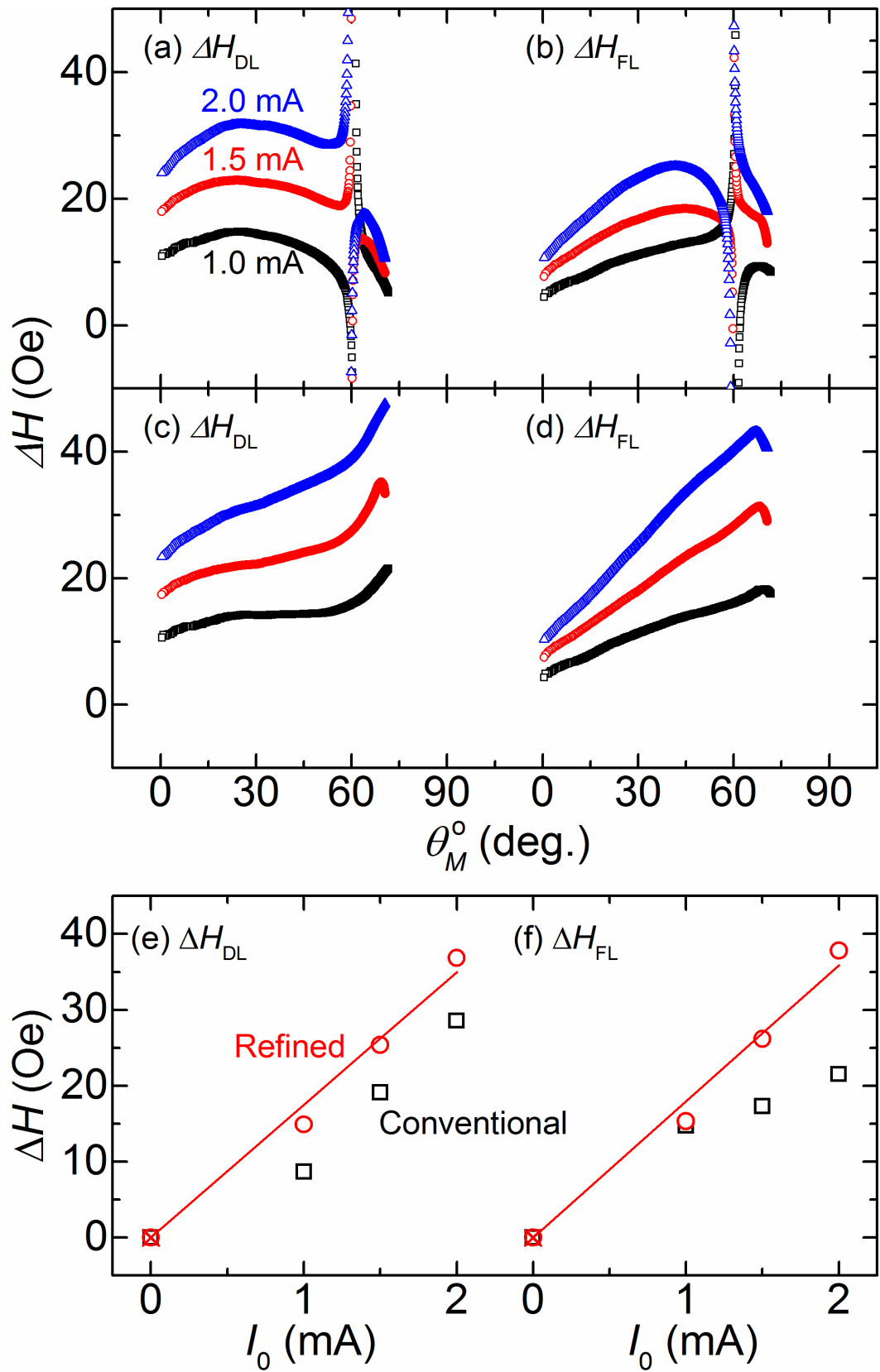

FIG. 6

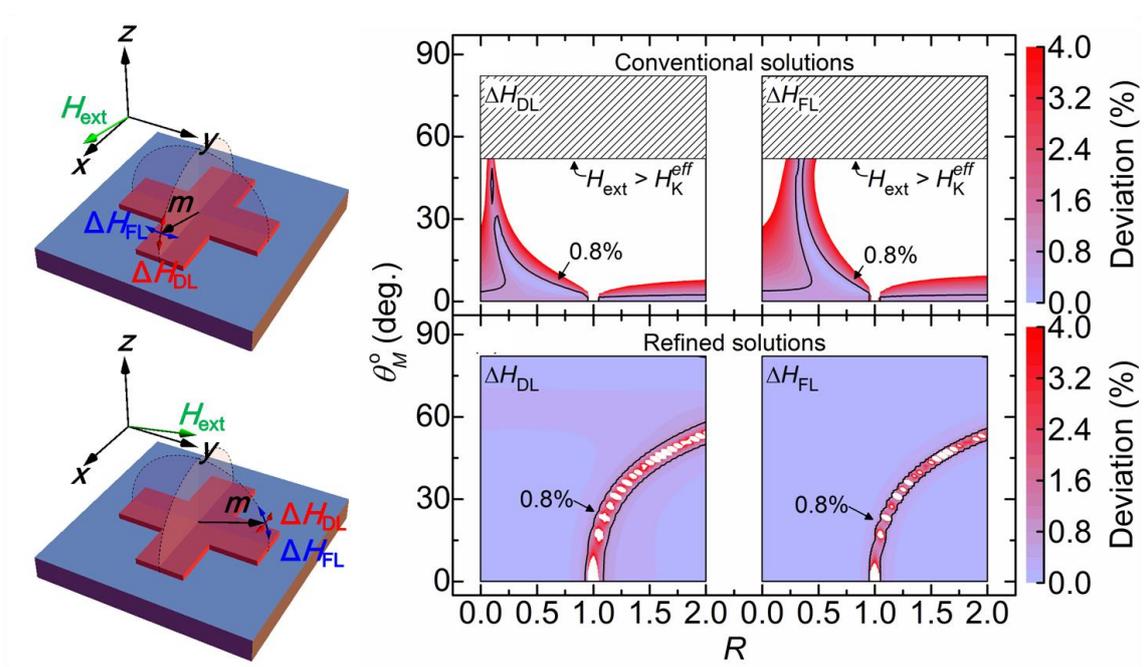

Graphical Abstract